\documentclass[12pt,preprintnumbers,nofootinbib,letterpaper]{article}
\pdfoutput=1
\usepackage{amsmath,amssymb,amsfonts,cite,xcolor,epsf,epsfig,epstopdf,graphics,booktabs,graphicx,mathtools,subcaption,physics,verbatim}
\def\thefootnote{*\arabic{footnote}}
\definecolor{ultramarine}{rgb}{0.07, 0.04, 0.56}
\definecolor{cadmiumgreen}{rgb}{0.0, 0.42, 0.24}
\definecolor{indigo(dye)}{rgb}{0.0, 0.25, 0.42}
\usepackage[linktocpage=true,breaklinks]{hyperref}
\hypersetup{
	colorlinks=true,
	citecolor=ultramarine,
	linkcolor=cadmiumgreen,
	urlcolor=indigo(dye),
}

\usepackage[utf8]{inputenc}

\numberwithin{equation}{section}

\usepackage{array}
\newcolumntype{P}[1]{>{\centering\arraybackslash}p{#1}}

\usepackage[shortlabels]{enumitem}

\usepackage{dcolumn}
\newcolumntype{M}[1]{>{\centering\arraybackslash}m{#1}}
\newcolumntype{N}{@{}m{0pt}@{}}

\usepackage{amsmath}
\usepackage{ulem}

\usepackage[utf8]{inputenc}
\usepackage[T1]{fontenc}
\usepackage{lmodern}
\usepackage{geometry}
\usepackage{amsmath,amssymb,amsthm,mathtools}
\usepackage{bm}
\usepackage{microtype}
\usepackage{physics}

\newcommand{\Mpl}{M_{\rm Pl}}

\newcommand{\D}{{\rm d}}

\newcommand{\be}{\begin{equation}}  
\newcommand{\ee}{\end{equation}}

\usepackage{tikz}

\begin{document}
	

\begin{center}
		
\def\thefootnote{\fnsymbol{footnote}}
		
\vspace*{1.5cm}
{\Large {\bf Covariant scalar-tensor theories beyond second derivatives}}
\\[1cm]
		
{Mohammad Ali Gorji$^{1}$, Pavel Petrov$^{1}$, Karim Noui$^{2}$}
\\[.7cm]

{\small\textit{$^{1}$Cosmology, Gravity, and Astroparticle Physics Group, Center for Theoretical Physics of the Universe, Institute for Basic Science (IBS), Daejeon, 34126, Korea}}\\
\vspace{0.25cm}
{\small\textit{$^{2}$Universit\'e Paris-Saclay, CNRS/IN2P3, IJCLab, 91405 Orsay, France}}
		
\end{center}
	
	
	
\begin{abstract}
We propose a covariant, gauge-independent construction of foliation-based scalar-tensor theories, yielding diffeomorphism-invariant operators involving only  gradients on the  hypersurfaces where the scalar field is constant, assumed to be spacelike. This defines a basis of independent invariants up to four derivatives of $\phi$, including the first nontrivial parity-odd pseudoscalar at this order, with a straightforward extension to higher derivatives. Our framework goes beyond degenerate higher-order scalar-tensor (DHOST) theories and provides a nonlinear extension of U-DHOST (where $\nabla_\mu\phi$ is supposed to be timelike) directly in covariant form, without using unitary gauge as a starting point or imposing degeneracy a priori. After minimal coupling to gravity, we analyze the theory through its Hamiltonian constraint structure and linear cosmological perturbations about an FLRW background, and show that it propagates three physical degrees of freedom.
\end{abstract}

	
\hrule
\def\thefootnote{\arabic{footnote}}
\setcounter{footnote}{0}


	
\newpage

\section{Introduction}\label{sec:introduction}

A wide class of modified gravity and early universe effective theories can be viewed as introducing, explicitly or dynamically, a preferred notion of time. Equivalently, spacetime is endowed with a distinguished foliation by three-dimensional hypersurfaces, which serves as the stage on which the extra gravitational dynamics is structured. This idea is familiar from several directions: Lorentz-violating gravity models where the foliation is fundamental, cosmological effective field theories where the time-slicing is fixed by the background evolution, and scalar-tensor theories where a scalar field provides a natural clock. Adopting a foliation-based description is useful because it makes transparent which operators are genuinely ``spatial'' on the slices and how an additional scalar mode can arise while still preserving the two tensor polarizations of general relativity.

This foliation-based viewpoint is often formulated in the language of spatially covariant gravity \cite{Gao:2014soa,Gao:2014fra}. In practice one describes the dynamics in terms of geometric data intrinsic to the spatial slices and their embedding in spacetime, rather than in manifestly four-dimensional form. This provides a common framework that encompasses Ho\v{r}ava-Lifshitz gravity \cite{Horava:2009uw,Blas:2009qj,Mukohyama:2010xz,Frusciante:2015maa} and the effective field theory of cosmology \cite{Arkani-Hamed:2003pdi,Cheung:2007st,Creminelli:2008wc,Gubitosi:2012hu,Bloomfield:2012ff,Gleyzes:2013ooa,Gleyzes:2014rba,Frusciante:2019xia}, among many others. A generic feature of such constructions is that once the time-slicing is singled out, the full spacetime diffeomorphism symmetry is reduced to diffeomorphisms that preserve the foliation. As a result, the constraint structure differs from that of general relativity, and in broad classes of models one typically finds two tensor modes accompanied by one additional scalar degree of freedom \cite{Gao:2014soa,Gao:2014fra}.

A complementary route starts from a fully spacetime-covariant scalar-tensor theory in which the foliation is defined by the level surfaces of a scalar field $\phi$. When $\nabla_\mu\phi$ is timelike, one can locally choose coordinates adapted to the foliation, so that $\phi$ plays the role of a clock and the resulting description makes contact with foliation-based operator languages. Conversely, many foliation-based actions can be expressed in a covariant form by introducing $\phi$ as a St\"uckelberg field \cite{Arkani-Hamed:2003pdi,Cheung:2007st}. This correspondence is powerful but also subtle: the covariant expression may display higher derivatives even when the foliation-based formulation involves at most second time derivatives, so that additional variables appearing in a covariant rewriting need not correspond to additional propagating degrees of freedom \cite{Blas:2010hb,Gleyzes:2013ooa,DeFelice:2018ewo,DeFelice:2021hps,DeFelice:2022xvq}.

Within covariant scalar-tensor theories, DHOST theories provide a systematic classification of higher-derivative Lagrangians that nevertheless propagate only one scalar degree of freedom (in addition to two tensor modes) thanks to covariant degeneracy conditions \cite{Langlois:2015cwa,BenAchour:2016cay,BenAchour:2016fzp}. More recently, the unitary gauge DHOST (U-DHOST) extension has clarified that degeneracy can be realized in a foliation-adapted manner: a covariant rewriting may appear to contain an extra mode, but this mode is non-propagating and instead obeys an elliptic (``shadowy''/instantaneous) equation on each hypersurface, when $\nabla_\mu\phi$ is supposed to be timelike. In this sense, U-DHOST can be regarded as a distinguished subclass of foliation-based theories characterized by a degeneracy property tied to the foliation \cite{DeFelice:2018ewo,DeFelice:2021hps,DeFelice:2022xvq}.

The goal of the present work is different in emphasis. Rather than imposing degeneracy conditions from the outset or using a foliation-adapted gauge as a starting point, we develop a unitary-gauge-independent covariant construction of foliation-adapted operators: diffeomorphism-invariant scalars built from $\phi$ and its derivatives that are guaranteed to reduce to purely spatial operators on the $\phi=\mathrm{const}$ hypersurfaces. Our strategy is geometric. We organize covariant building blocks using Gram determinants (equivalently, wedge products) of gradients of scalars, which automatically isolates combinations in which components along the foliation normal cancel. This yields a compact basis of independent invariants up to four derivatives of $\phi$ in the scalar-gradient sector, including the first nontrivial parity-odd pseudoscalar at this order, and it extends straightforwardly to higher derivatives.

The rest of the paper is organized as follows. In Sec.~\ref{sec-construction} we develop a covariant, gauge-independent construction of foliation-adapted operators and derive a compact basis of independent invariants up to four derivatives of $\phi$, including the leading parity-odd pseudoscalar. In Sec.~\ref{sec-model} we introduce a minimally coupled model defined by an arbitrary function of these invariants and we compare it with the DHOST and U-DHOST theories. In Sec.~\ref{sec:dof}, we study the dynamics of the model through the Hamiltonian constraint structure and linear cosmological perturbations around a spatially flat Friedmann-Lema\^{i}tre-Robertson-Walker (FLRW) background. We show that the higher-derivative invariants vanish on the homogeneous FLRW background and contribute only to fluctuations, so that the theory propagates three degrees of freedom, two tensor modes and one scalar mode. Sec.~\ref{sec:Summary} is devoted to the summary and conclusions. 

\section{Construction}\label{sec-construction}

We now turn to the construction of a covariant operator basis adapted to the foliation defined by the scalar field $\phi$ (assuming $\phi_\mu\equiv\nabla_\mu\phi$ is timelike). The strategy is to proceed order by order in derivatives of $\phi$, and at each order identify diffeomorphism scalars that encode only the intrinsic derivative data on the hypersurfaces $\phi=\mathrm{const}$. In practice, we will first construct the lowest-order ``safe'' invariants and then use them as seeds for the higher-derivative building blocks in the subsequent subsections.

We consider a higher-derivative scalar-tensor theory built from $\phi$ and $g_{\mu\nu}$, subject to the following guiding principles:
\begin{enumerate}
	\renewcommand{\theenumi}{\Roman{enumi}}
	\renewcommand{\labelenumi}{\theenumi.}
	\item \textbf{Connection free:} building blocks are taken to be scalars and (exterior) derivatives of scalars, so that Christoffel symbols do not appear explicitly.
	\item \textbf{Spatial foliation:} admissible operators must probe only directions tangent to the hypersurfaces $\phi=\mathrm{const}$, i.e.\ they should be insensitive to components parallel to $\phi_\mu$. In our construction this is implemented by forming Gram-determinant (equivalently, wedge-product) combinations involving $\D\phi$, which automatically remove any part proportional to $\D\phi$.\footnote{More precisely, the tangential information carried by a covector $\omega$ on $\phi=\mathrm{const}$ is captured by the equivalence class $[\omega]\in T_x^*M/\mathrm{span}\{\D\phi\}$, i.e.\ $\omega\sim\omega+\lambda\,\D\phi$. Wedge products with $\D\phi$ defines an action on this class, since $\D\phi\wedge(\omega+\lambda\,\D\phi)=\D\phi\wedge\omega$. In components, this is the invariance under $\omega_\mu\to \omega_\mu+\lambda\,\phi_\mu$.}
\end{enumerate}

\subsection{Up to second order}

In this subsection we classify independent scalar building blocks containing up to second derivatives of $\phi$. 

At zeroth order in $\phi_{\mu\nu}=\nabla_\nu\nabla_\mu\phi$, the independent scalar invariants are
\begin{align}\label{X-def}
	\phi,
	\qquad
	X \equiv g^{\mu\nu}\phi_\mu\phi_\nu.
\end{align}
To extract intrinsic derivative information on $\phi=\mathrm{const}$ one needs at least two independent gradients. At the next derivative order this  becomes possible through $\D\phi$ and $\D X$ (since $X_\mu=\nabla_\mu X$ already contains $\phi_{\mu\nu}$), and the natural foliation-intrinsic object is their wedge product.

At first order in $\phi_{\mu\nu}$, a natural scalar built without introducing explicit connection coefficients is
\begin{align}\label{Y-def}
	Y \equiv \phi^\mu X_\mu.
\end{align}
We will use $Y$ as a convenient intermediate quantity, but it is not foliation-intrinsic by itself, since it is sensitive to the component of $X_\mu$ parallel to $\phi_\mu$.

At second order in $\phi_{\mu\nu}$, there are two independent scalar contractions constructed from $X_\mu$ and $\phi_\mu$,
\begin{align}\label{Z-def}
	Y^2,
	\qquad
	Z \equiv X^\mu X_\mu.
\end{align}
However, the foliation-intrinsic information contained in $\D X$ is more naturally packaged in the two-form
\begin{align}
{\cal F}_{\mu\nu}\equiv \phi_\mu X_\nu-\phi_\nu X_\mu,
\end{align}
which is precisely the component expression of $\D\phi\wedge \D X$ (and is therefore reminiscent of an electromagnetic field strength built from scalar ``potentials''). At this derivative order, the associated scalar invariant is uniquely obtained by taking the squared norm of this two-form, which yields the distinguished combination\footnote{The squared-norm representation \eqref{eq:B-def} can also be written using the Levi-Civita tensor as
	\begin{align}\label{B-wedge}
		B=\frac12\,{\cal F}_{\mu\nu}{\cal F}^{\mu\nu}
		=-\frac12\,\epsilon^{\mu\nu\alpha\beta}\epsilon_{\mu\nu\rho\sigma}\,
		\phi_\alpha X_\beta\,\phi^\rho X^\sigma,
	\end{align}
	which makes it explicit that $B$ is built from the wedge product $\D\phi\wedge \D X$ and therefore satisfies Principle~II by construction. }
\begin{align}\label{eq:B-def}
B \equiv \frac12\,{\cal F}_{\mu\nu}{\cal F}^{\mu\nu}=XZ-Y^2 \,.
\end{align}
We now show that $B$ is foliation-spatial.

Assuming $\phi_\mu$ is timelike, define the unit normal to the $\phi=\mathrm{const}$ hypersurfaces and the corresponding projector,
\begin{align}\label{eq:n-h-def}
	n_\mu \equiv -\frac{\phi_\mu}{\sqrt{-X}},
	\qquad
	h_{\mu\nu}\equiv g_{\mu\nu}+n_\mu n_\nu,
\end{align}
so that $n_\mu n^\mu=-1$ and $h^\mu{}_\nu n^\nu=0$. Using $n_\mu\propto \phi_\mu$ and $X_\mu=\partial_\mu X$, one finds
\begin{align}\label{B-spatial}
	B = X\,h^{\mu\nu}X_\mu X_\nu.
\end{align}
Equation \eqref{B-spatial} makes it manifest that $B$ depends only on the gradient of $X$ tangent to the $\phi=\mathrm{const}$ hypersurfaces. In particular, as we will see, in coordinates adapted to the foliation (e.g.\ unitary gauge $\phi=t$), \eqref{B-spatial} reduces to a purely spatial expression and hence contains no higher derivatives along the normal/time direction.

In the remainder of the paper we will often use the equivalent representation in terms of $n_\mu$ and $h_{\mu\nu}$ (as in \eqref{B-spatial}), since it provides a convenient bookkeeping device for identifying independent invariants at higher derivative order. Note that this does not amount to gauge fixing: it is simply a covariant rewriting of the same wedge/Gram-determinant construction.

\subsection{Up to third order}\label{sec:third}

In this subsection we extend the construction to include up to third derivatives of $\phi$, while keeping the same guiding principles as in the previous subsection, i.e., conditions I and II.

We adopt a recursive strategy: first construct safe (foliation-spatial) scalars, then increase derivative order by taking gradients of those scalars. Because the gradient of a scalar is a covector (one-form), this preserves the ``no explicit connection'' property at the level of building blocks. Moreover, if the scalar is already foliation-spatial, then its projected gradient generates foliation-spatial invariants at the next derivative order.

Allowing third derivatives means we can now use the covector
\begin{align}\label{def:Bmu}
	B_\mu = Z\,X_\mu + X\,Z_\mu - 2Y\,Y_\mu \,,
\end{align}
which contains up to third derivatives of $\phi$. This relation should be understood as the identity $B_\mu=\nabla_\mu B$ following from $B=XZ-Y^2$. In our construction the relevant ``safe'' object is its tangential projection $D_\mu B\equiv h_\mu{}^{\nu}\nabla_\nu B$. Treating $Y_\mu$ and $Z_\mu$ as independent building blocks would amount to reproducing the same cancellations less systematically.

Using \eqref{def:Bmu} as a safe building block, we define the two independent third-order invariants
\begin{equation}\label{def:C1C2-proj}
\begin{split}
		C^{[1]} \equiv X\,h^{\mu\nu}X_\mu B_\nu \,,
		\qquad
		C^{[2]} \equiv X\,h^{\mu\nu}B_\mu B_\nu \,.
\end{split}
\end{equation}

These are the unique (up to functions of lower-order invariants) scalars built from the new covector $B_\mu$ without introducing second derivatives of scalars as basic building blocks.

One may wonder if the two invariants \eqref{def:C1C2-proj} exhaust the independent scalars at third derivative order in the present construction. Indeed, any scalar $S$ built from the lower-order invariants $(\phi,X,B)$ has a gradient $S_\mu = S_{,\phi}\,\phi_\mu + S_{,X}\,X_\mu + S_{,B}\,B_\mu$. Since $D_\mu \phi \equiv h_\mu{}^\nu\nabla_\nu\phi = 0$ identically, it follows $D_\mu S= S_{,X}\, D_\mu X + S_{,B}\,D_\mu B$, showing that the space of available spatial covectors is two-dimensional, spanned by $\{D_\mu X, D_\mu B\}$.

\subsection{Up to fourth order}\label{sec:fourth}

In this subsection, we extend the construction
up to fourth derivatives of $\phi$.
We can now use the covectors obtained by differentiating the third-order scalars
\begin{align}\label{def:C1muC2mu} 
C^{[1]}_\mu \,,
\qquad
C^{[2]}_\mu \,.
\end{align}
Their projections tangent to the foliation involve up to fourth derivatives of $\phi$ and do not explicitly include the connection.

At fourth derivative order the relevant spatial covectors are therefore 
\begin{align}
D_\mu X,\qquad D_\mu B,\qquad D_\mu C^{[1]},\quad  D_\mu C^{[2]}.
\end{align}
However, three contractions are already fixed by lower-order invariants, 
\begin{eqnarray}
h^{\mu\nu}D_\mu X D_\nu{X} = \frac{B}{X} \, , \quad h^{\mu\nu}D_\mu X D_\nu{B}=\frac{C^{[1]}}{X} \, , \quad h^{\mu\nu}D_\mu B D_\nu{B} =\frac{C^{[2]}}{X} \, .
\end{eqnarray}
Hence, the genuinely new information at fourth order is encoded in the six remaining structures involving $D_\mu C^{[1,2]}$. A convenient choice of independent fourth-order invariants is
\begin{equation}\label{def:D123-nonorth}
\begin{aligned}
D^{[1]} &\equiv X\,h^{\mu\nu} X_\mu\, C^{[1]}_\nu \,, 
&\qquad
D^{[2]} &\equiv X\,h^{\mu\nu} B_\mu\, C^{[1]}_\nu \,, 
&\qquad
D^{[3]} &\equiv X\,h^{\mu\nu} C^{[1]}_\mu C^{[1]}_\nu \,, 
\\
E^{[1]} &\equiv X\,h^{\mu\nu} X_\mu\, C^{[2]}_\nu \,, 
&\qquad
E^{[2]} &\equiv X\,h^{\mu\nu} B_\mu\, C^{[2]}_\nu \,, 
&\qquad
E^{[3]} &\equiv X\,h^{\mu\nu} C^{[1]}_\mu C^{[2]}_\nu \,.
\end{aligned}
\end{equation}
The additional contraction $E^{[4]} \equiv X\,h^{\mu\nu} C^{[2]}_\mu C^{[2]}_\nu$ is redundant due to a dimension-dependent identity; the proof is presented in Appendix~\ref{app:Independence of the Basis}.  These six quantities may therefore be taken as a basis of independent building blocks at fourth-derivative order within the scalar-gradient sector, away from degenerate configurations in which $D_\mu C^{[1]}$ and/or $D_\mu C^{[2]}$ lie in the span of $\{D_\mu X, D_\nu B\}$.

At fourth order, a genuinely parity-odd scalar becomes possible for the first time in the scalar-gradient sector. The reason is purely geometric: a parity-odd scalar in four dimensions requires a single Levi-Civita tensor contracted with four independent covectors, schematically $ \phi_\mu \epsilon^{\mu\nu\rho\sigma}\,D_{\nu}X\,D_{\rho}B\,D_{\sigma} C^{[1]}$. Up to third order we only had three independent covectors $\{\phi_{\mu},D_{\nu} X,D_{\rho} B\}$, so any such contraction vanished. At fourth order, two new covectors emerge $D_{\mu} C^{[1,2]}$. Either of their components orthogonal to the $(D_{\mu}X,D_{\nu}B)$ plane provides the needed fourth direction, and we may define
\begin{align}\label{def:Dodd}
D^{[\mathrm{odd}]}\equiv n_\mu \epsilon^{\mu\nu\rho\sigma}\,D_{\nu}X\,D_{\rho}B\,D_{\sigma} C^{[1]}
,
\end{align}
where $n_\mu$ is defined in Eq.~\eqref{eq:n-h-def}. Indeed, $(D^{[\mathrm{odd}]})^2$ is not an independent quantity:
\begin{align}\label{eq:Dodd2}
\big(D^{[\mathrm{odd}]}\big)^2 
   \;&=\; \frac{1}{X^3}
\begin{vmatrix}
B & C^{[1]} & D^{[1]}\\
C^{[1]} & C^{[2]} & D^{[2]}\\
D^{[1]} & D^{[2]} & D^{[3]}
\end{vmatrix}\,,
\end{align}
which shows that $D^{[\mathrm{odd}]}$ captures only the \textit{parity} of the field configuration, while its absolute value does not provide additional independent information. Consequently, any other parity-odd pseudoscalar obtained by including spatial gradients, such as $dC^{[2]}$, is redundant on the generic nondegenerate branch; see Appendix~\ref{app:Independence of the Basis} for a proof. Next, we can define the sign/orientation variable
\begin{align}\label{eq:def-calE}
{\cal E}^{[\mathrm{odd}]} \equiv \frac{D^{[\mathrm{odd}]}}{\sqrt{\big(D^{[\mathrm{odd}]}\big)^2}}\,,\quad \big({\cal E}^{[\mathrm{odd}]}\big)^2 = 1.   
\end{align}

In summary, on the generic nondegenerate branch, the building blocks involving up to fourth derivatives of the scalar field are
\begin{align}
\label{eq:basis}
\boxed{
\phi \,,\quad 
X \,,\quad 
B \,,\quad 
C^{[1,2]} \,,\quad 
D^{[1,2,3]} \,,\quad 
E^{[1,2,3]}
\,\quad
{\cal E}^{[\mathrm{odd}]} 
}.
\end{align}
These quantities form a basis for the algebra of elementary Lagrangians up to fourth order in derivatives. The proof is given in Appendix~\ref{app:Independence of the Basis}.

\section{The model}\label{sec-model}

In this section we minimally couple the scalar sector constructed above to gravity. Since each element of the basis \eqref{eq:basis} is, by construction, a diffeomorphism scalar that is foliation-spatial on $\phi=\mathrm{const}$, the most general parity-even Lagrangian in this sector can depend on these building blocks through an arbitrary function. Although ${\cal E}^{[\mathrm{odd}]}$ is a convenient variable for tracking parity, it is nonanalytic at $D^{[\mathrm{odd}]}=0$. We therefore use $D^{[\mathrm{odd}]}$ instead, which is defined in \eqref{def:Dodd}. In that case, $\big(D^{[\mathrm{odd}]}\big)^2$ is not an independent invariant but is fixed by the parity-even basis variables (see \eqref{eq:Dodd2}). As a result, any local dependence on $D^{[\mathrm{odd}]}$ can be reduced, without loss of generality, to a parity-even part plus a parity-odd part proportional to $D^{[\mathrm{odd}]}$. Thus we may take the action to be at most linear in $D^{[\mathrm{odd}]}$. We then consider the minimally coupled action
\begin{align}\label{eq:action}
	\begin{split}
		S &= \int \mathrm{d}^4x \, \sqrt{-g} \Bigg[
		\frac{M_{\rm Pl}^2}{2} R
		+ P\!\left(\phi, X, B, C^{[1,2]}, D^{[1,2,3]}, E^{[1,2,3]}\right)
		\\
		&\hspace{4cm}
		+ \tilde{P}\!\left(\phi, X, B, C^{[1,2]}, D^{[1,2,3]}, E^{[1,2,3]}\right)\,D^{[\mathrm{odd}]}
		\Bigg] \,,
	\end{split}
\end{align}
where the building blocks are defined in \eqref{X-def}, \eqref{eq:B-def}, \eqref{def:C1C2-proj}, \eqref{def:D123-nonorth}, and  \eqref{def:Dodd}, with 
$\phi$ allowed as an additional zeroth-order scalar.  Here $P$ and $\tilde{P}$ are arbitrary functions of the parity-even building blocks in \eqref{eq:basis}. By construction, these arguments are diffeomorphism scalars built from $\phi$ and its derivatives and are adapted to the foliation $\phi=\mathrm{const}$ in the sense explained in Sec.~\ref{sec-construction}. 

It is useful to relate the scalar-gradient building blocks introduced above to the standard DHOST basis of operators quadratic in $\phi_{\mu\nu}$, denoted $L_i^{(2)}$ in Ref.~\cite{BenAchour:2016fzp}. In particular,
\begin{align}
	L_4^{(2)} &\equiv \phi^\mu \phi_{\mu\rho}\,\phi^{\rho\nu}\phi_\nu \,,
	\qquad
	L_5^{(2)} \equiv (\phi^\mu\phi_{\mu\nu}\phi^\nu)^2 \,.
\end{align}
Using $X_\mu= 2\,\phi^\nu \phi_{\nu\mu}$, one finds
\begin{align}
	Y &= \phi^\mu X_\mu = 2\,\phi^\mu \phi_{\mu\nu}\phi^\nu \,,
	\qquad
	Z = X^\mu X_\mu = 4\,\phi^\mu \phi_{\mu\rho}\,\phi^{\rho\nu}\phi_\nu = 4\,L_4^{(2)} \,,
\end{align}
so that the second-derivative building block \eqref{eq:B-def} can be written as the specific quadratic combination\footnote{It is worth mentioning that in the singular limit of consistent higher-derivative conformal transformations studied in \cite{Babichev:2019twf,Babichev:2021bim,Takahashi:2021ttd,Babichev:2024eoh}, the allowed higher-derivative contributions up to second derivatives of the scalar can be organized in terms of the two dimensionless quantities $4L_4^{(2)}/X^3$ and $2\sqrt{L_5^{(2)}}/X^2$, which indeed contains $4\big(X\,L_4^{(2)}-L_5^{(2)}\big)/X^4$ as a special subset \cite{Gorji:2025ajb}.}
\begin{align}\label{eq:B-to-DHOST}
	B = XZ-Y^2 = 4\Big(X\,L_4^{(2)}-L_5^{(2)}\Big)\,.
\end{align}
This observation is helpful for comparison, but it should not be confused with the DHOST construction itself: DHOST theories are singled out by degeneracy conditions on the coefficients of $L_i^{(2)}$, with $i=1,\cdots,5$ \cite{BenAchour:2016fzp}, and the isolated term \eqref{eq:B-to-DHOST} does not, by itself, satisfy these conditions. In this sense, allowing for a general function of $B$ already goes beyond the standard DHOST classification unless one imposes additional restrictions.

Moreover, our third-derivative invariants $C^{[1]}$ and $C^{[2]}$ are constructed from the projected gradient $D_\mu B$ and therefore involve derivatives of $\phi_{\mu\nu}$ (i.e.\ genuine third derivatives of $\phi$). They thus lie outside the usual DHOST operator basis, where ``cubic'' refers to terms cubic in $\phi_{\mu\nu}$ without derivatives acting on $\phi_{\mu\nu}$ \cite{BenAchour:2016fzp}. This is precisely the sense in which our covariant, foliation-adapted basis extends the familiar DHOST building blocks.

Finally, it is worth placing these structures in the broader landscape of foliation-based theories. The combination \eqref{eq:B-to-DHOST} (and, more generally, operators that are manifestly spatial in unitary gauge) naturally appears in the context of unitary gauge degenerate extensions (U-DHOST), in which the degeneracy is tied to the foliation and an apparent extra covariant mode can be ``shadowy'' rather than propagating \cite{DeFelice:2018ewo,DeFelice:2021hps,DeFelice:2022xvq}.

\section{Degrees of freedom}\label{sec:dof}

To determine the number of physical degrees of freedom propagated by the theory, it is convenient to work in a foliation-adapted coordinate system in which the scalar field labels the hypersurfaces, i.e.\ in unitary gauge
\begin{align}\label{eq:unitary-gauge}
\phi=t \, .
\end{align}
Although our construction is fully covariant, unitary gauge makes the foliation-spatial character of the building blocks completely explicit and provides a natural starting point for both the Hamiltonian analysis and the cosmological perturbation theory.

\subsection{Hamiltonian analysis in the unitary gauge}
Using the standard ADM decomposition of the metric,
\begin{eqnarray}\label{eq:metric-ADM}
	{\D}s^2 = - N^2 {\D}t^2 + h_{ij}({\D}x^i + N^i {\D}t)({\D}x^j+N^j {\D}t) \, ,
\end{eqnarray} 
where $N$ is the lapse function, $N^i$ the shift vector, and $h_{ij}$ the spatial metric, one has
\begin{align}
	X=-\frac{1}{N^2} \,,
\end{align}
and the higher-derivative invariants reduce to purely spatial combinations on the constant-$t$ slices,
\begin{align}\label{eq:BB-unitary}
\begin{aligned}
	B&=X\,h^{ij}\,\partial_i X\,\partial_j X \,,
	&&C^{[1]}=X\,h^{ij}\,\partial_i X\,\partial_j B \,,
	&&C^{[2]}=X\,h^{ij}\,\partial_i B\,\partial_j B \,,
	\\
	D^{[1]}&=X\,h^{ij}\,\partial_i X\,\partial_j C^{[1]} \,,
	&&D^{[2]}=X\,h^{ij}\,\partial_i B\,\partial_j C^{[1]} \,,
	&&D^{[3]}=X\,h^{ij}\,\partial_i C^{[1]}\,\partial_j C^{[1]} \,,
	\\
	E^{[1]}&=X\,h^{ij}\,\partial_i X\,\partial_j C^{[2]} \,,
	&&E^{[2]}=X\,h^{ij}\,\partial_i B\,\partial_j C^{[2]} \,,
	&&E^{[3]}=X\,h^{ij}\,\partial_i C^{[1]}\,\partial_j C^{[2]} \,.
\end{aligned}
\end{align}
For the parity-odd structure, using $\epsilon^{0ijk}=\sqrt{-X}\,\epsilon^{ijk}=\epsilon^{ijk}/N$, we obtain
\begin{align}
	D^{[\mathrm{odd}]}
	=-\epsilon^{ijk}\,\partial_i X\,\partial_j B\,\partial_k C^{[1]} \,,
\end{align}
where $\epsilon^{ijk}$ is the Levi-Civita tensor on the spatial slice that is compatible with $h_{ij}$.

The key point is that, in unitary gauge, all these operators involve only spatial derivatives $\partial_i$ and the spatial metric $h_{ij}$, and hence contain no higher derivatives along the time direction. This makes it possible to analyze the constraint structure in a standard Hamiltonian framework and, independently, to study the propagating content through linear cosmological perturbations. We now turn to these two complementary analyses.

The Hamiltonian analysis is straightforward. Indeed, without going into the details, we begin with 
20 phase-space degrees of freedom, corresponding to the ten components of the metric and their conjugate momenta.

Since the Lagrangian does not contain $\dot N$ (despite the breaking of time-reparametrization symmetry), the theory admits four primary constraints, arising from the vanishing canonical momenta associated with the lapse and the shift vector. The requirement that these primary constraints remain stable under time evolution leads to four additional constraints: three corresponding to the vector (momentum) constraint and one additional constraint that replaces the Hamiltonian constraint. 
In general, if $P$ and $\tilde P$ are generic and are associated with no particular symmetries, the Dirac algorithm stops at this stage with a total of eight constraints: six first-class constraints and two second-class constraints. Therefore we immediately conclude that the theory admits $ (20- 2 \cdot 6 - 2) = 2 \cdot 3$ degrees of freedom in the phase space corresponding to the tensor modes and one scalar mode. In principle, additional constraints could arise, leading to fewer than three degrees of freedom.

Away from unitary gauge, i.e. in the fully covariant formulation, higher-order time derivatives may appear, and a Hamiltonian analysis may suggest the presence of additional degrees of freedom. However, as shown in \cite{DeFelice:2018ewo}, such apparently extra modes can, in appropriate cases, correspond to so-called ``shadowy'' modes rather than genuine propagating degrees of freedom.

\subsection{Cosmological perturbations}

In this subsection, we study the cosmological implications of our model. We show that the higher-derivative operators do not affect the homogeneous FLRW background and that their first nontrivial effects appear at the level of linear perturbations. We also confirm that the theory propagates three degrees of freedom.

Considering the spatially flat FLRW background and homogeneous scalar field
\begin{align}
{\bar N}= \bar{N}(t)\,, \qquad
\bar{N}_i = 0 \,, \qquad
\bar{h}_{ij} = a^2(t) \delta_{ij} \,, \qquad 
\bar{\phi} = \bar{\phi}(t) \,,
\end{align}
where $a(t)$ is the scale factor and $\bar{N}(t)$ is the background lapse function, we find
\begin{align}
	\bar{X} = - \dot{\bar{\phi}}^2 \,,
	\qquad
	\bar{Y} = - \dot{\bar{\phi}} \dot{\bar{X}} \,,
	\qquad
	\bar{Z} = - \dot{\bar{X}}^2 \,,
\end{align}
where a dot is defined as $\dot{}\equiv \tfrac{\D}{\bar{N}\D{t}}$. The above results show that
\begin{align}\label{eq:B-BG}
\bar{B} = \bar{C}^{[1,2]} = \bar{D}^{[1,2,3]} = \bar{E}^{[1,2,3]} = \bar{D}^{[\mathrm{odd}]} = 0 \,.
\end{align}

The Einstein equations give the first and second Friedmann equations as
\begin{align}\label{eq:Friedmann1}
	3 \Mpl^2 H^2 &= \bar{\rho} \,,
	\\ \label{eq:Friedmann2}
	\Mpl^2\left( 2 \dot{H} + 3 H^2 \right) &= - \bar{p} \,,
\end{align}
where $H=\dot{a}/a$ is the Hubble parameter and the background energy density and pressure are defined as
\begin{align}\label{eq:rho-BG}
\bar{\rho} &\equiv 2 \bar{X} \bar{P}_{,\bar{X}} - \bar{P} \,,
\qquad
\bar{p} \equiv \bar{P}\,.
\end{align}
Note that, since all higher-derivative quantities vanish on the cosmological background \eqref{eq:B-BG}, the equations of motion on an FLRW background are exactly the same as in the k-essence model \cite{Armendariz-Picon:1999hyi,Garriga:1999vw,Armendariz-Picon:2000ulo,Chen:2006nt}. Thus, the higher-derivative terms only affect perturbations. Clearly, all FLRW solutions that have been found for k-essence theory are also solutions of our theory.

To see the first nontrivial effects of the higher-derivative operators in cosmology, we therefore turn to perturbations. We use the time diffeomorphism to work in comoving gauge,
\begin{align}\label{eq:unitaryG}
	\phi(t,x)=\bar{\phi}(t)\,,
\end{align}
so that the scalar field has no fluctuations. One may further choose $\bar{\phi}(t)=t$ and $\bar{N} = 1$ (which coincides with the unitary gauge defined in \eqref{eq:unitary-gauge}), in which case $\bar{X}=-1$. For later convenience, however, we keep $\bar{\phi}(t)$ and $\bar{X}$ arbitrary. We then use the scalar part of the spatial diffeomorphisms to fix the metric perturbations as
\begin{align}\label{eq:metric-pert}
	N = \bar{N}\left(1+\alpha\right)\,, \qquad
	N_i = \bar{N}\,\partial_i \chi\,, \qquad
	h_{ij} = a^2 \left(1 + 2 \zeta \right) e^{\gamma_{ij}}\,,
\end{align}
where $\alpha$, $\chi$, and $\zeta$ are scalar perturbations, and $\gamma_{ij}$ is a transverse and traceless tensor perturbation.

Using \eqref{eq:unitaryG} and \eqref{eq:metric-pert}, and expanding up to second order in scalar perturbations, we obtain
\begin{align}\label{eq:X-upTo2nd}
	X = \bar{X}\left(1-2\alpha+3\alpha^2\right) \;+\;\mathcal{O}(\epsilon_s^3)\,,
\end{align}
where $\epsilon_s$ denotes the amplitude of scalar perturbations (here represented by $\alpha$). It follows that $\partial_i X=\mathcal{O}(\epsilon_s)+\cdots$, where $\cdots$ denotes higher-order corrections starting at $\mathcal{O}(\epsilon_s^2)$. Using the unitary gauge expressions \eqref{eq:BB-unitary}, one then immediately sees that $B$ starts at quadratic order, since it contains two spatial gradients of $X$. Indeed, keeping only the leading contribution one finds
\begin{align}\label{eq:B-upTo2nd}
B =  4\,\bar{X}^3\,\bar{h}^{ij}\,\partial_i\alpha\,\partial_j\alpha
	\;+\;\mathcal{O}(\epsilon_s^3)\,.
\end{align}
The remaining building blocks are higher order in perturbations. From \eqref{eq:BB-unitary} and the scalings $\partial_i X=\mathcal{O}(\epsilon_s)$ and $\partial_i B=\mathcal{O}(\epsilon_s^2)$, we obtain the parametric orders
\begin{align}
	C^{[1]}=\mathcal{O}(\epsilon_s^3),\hspace{.4cm}
	C^{[2]}=\mathcal{O}(\epsilon_s^4),\hspace{.4cm}
	D^{[1]}=\mathcal{O}(\epsilon_s^4),\hspace{.4cm}
	D^{[2]}=\mathcal{O}(\epsilon_s^5),\hspace{.4cm}
	D^{[3]}=\mathcal{O}(\epsilon_s^6),
\end{align}
and similarly
\begin{align}
	E^{[1]}=\mathcal{O}(\epsilon_s^5),\hspace{.4cm}
	E^{[2]}=\mathcal{O}(\epsilon_s^6),\hspace{.4cm}
	E^{[3]}=\mathcal{O}(\epsilon_s^7),\hspace{.4cm}
	D^{[\mathrm{odd}]}=\mathcal{O}(\epsilon_s^6).
\end{align}
Therefore, up to quadratic order in scalar perturbations, the only higher-derivative invariant that contributes is $B$. The next correction arises at cubic order through $C^{[1]}$, while the remaining invariants $C^{[2]}$, $D^{[1,2,3]}$, $E^{[1,2,3]}$, and $D^{[\mathrm{odd}]}$ start contributing only at quartic order and beyond. 

Substituting \eqref{eq:unitaryG} and \eqref{eq:metric-pert} into the action \eqref{eq:action}, then expanding up to the second order in perturbations and performing some integration by parts,  we find for the scalar perturbations
\begin{align}\label{eq:action-quadratic-1}
	S^{(2)} =\int \D{t}\D^3x {\bar N} a^3 
	\big[
	{\cal L}^{(2)}_{\rm E.H.} + {\cal L}^{(2)}_{\rm P}
	\big] \,,
\end{align}
where we have defined quadratic Lagrangian densities
\begin{align}\label{LEH}
	{\cal L}^{(2)}_{\rm E.H.} &\equiv \Mpl^2
	\bigg[ 3 \left(3\dot{H}+\frac{9}{2} H^2 \right) \zeta^2
	-3 \dot{\zeta}^2 
	+ 3 H \alpha \left( 2\dot{\zeta}+3H\zeta - H\alpha\right)
	\\
	&\hspace{2cm}
	+\frac{\left(\partial\zeta\right)^2}{a^2} 
	+ 2\alpha \frac{\partial^2\zeta }{a^2}
	+ 2 \left( \dot{\zeta} - H \alpha \right) \frac{\partial^2\chi}{a^2}
	\bigg] \,,
	\\
	\begin{split}\label{LM}
	{\cal L}^{(2)}_{\rm P} &\equiv 
	\frac{\epsilon}{c_s^2} \Mpl^2 H^2 
	\alpha \left(\alpha - 6 c_s^2 \zeta \right)
	+ \frac{3}{2} \bar{P} \zeta (3 \zeta + 2 \alpha) 
	- 4 \bar{X}^3 \bar{P}_{,B}  \alpha \frac{\partial^2 \alpha}{a^2}
		\,,
	\end{split}
\end{align}
with
\begin{align}
\epsilon &\equiv \frac{\bar{X}\bar{P}_{,\bar{X}}}{\Mpl^2 H^2} \,,
\qquad
c_s^2 \equiv \frac{\bar{p}_{,\bar X}}{\bar{\rho}_{,\bar X}} \,.
\end{align}
Note that the parity-odd invariant $D^{[\mathrm{odd}]}$ starts contributing only at sixth order in scalar perturbations and, therefore, the term proportional to $\tilde{P}$ in \eqref{eq:action} does not contribute to the quadratic action for linear perturbations.

Integrating out the non-dynamical fields $\alpha$ and $\chi$, using background equations, and going to Fourier space, we find
\begin{align}\label{eq:action-S2}
S^{(2)}_{\rm S} =\Mpl^2\,\int \frac{\D^3k}{(2\pi)^3} \D{t}\, {\bar N} a^3 
\left[  
\left(
\frac{\epsilon}{c_s^2} + \beta \frac{k^2}{a^2H^2}
\right) \dot{\zeta}_k^2
- \epsilon \frac{k^2}{a^2} \zeta_k^2
\right] \,,
\end{align}
where we have defined the dimensionless quantity
\begin{align}
\beta \equiv \frac{4\bar{X}^3\bar{P}_{,\bar B}}{\Mpl^2} \,.
\end{align}
The parameter $\beta$ encodes all effects of the higher-derivative terms on linear cosmological perturbations such that in the limit $\beta = 0$, one recovers the well-known k-essence result \cite{Garriga:1999vw,Chen:2006nt}.

At low physical momenta, $c_s k/a\ll H\sqrt{\epsilon/\beta}$ (superhorizon modes when $\sqrt{\epsilon/\beta}=\mathcal{O}(1)$), one recovers the usual linear dispersion relation $\omega^2\simeq c_s^2 k^2/a^2$ \cite{Garriga:1999vw,Chen:2006nt}. By contrast, for $c_s k/a\gg H\sqrt{\epsilon/\beta}$ (subhorizon modes when $\sqrt{\epsilon/\beta}=\mathcal{O}(1)$), the $\beta$-term dominates the kinetic coefficient and the frequency saturates to a $k$-independent value, $\omega^2\simeq \epsilon H^2/\beta$, so that the group velocity tends to zero. This UV freezing is a distinctive feature of the $k^2\dot\zeta_k^2$ correction: it enhances the effective kinetic weight of short-wavelength fluctuations while leaving their leading spatial-gradient structure unchanged, i.e.\ it does not by itself generate an additional stabilizing gradient term. This point becomes important in stealth-like regimes where $\epsilon\to 0$ (for instance when $\bar{P}_{,{\bar X}}\to 0$ while $\epsilon/c_s^2\propto\bar{P}_{,{\bar{X}\bar{X}}}$ stays finite): the standard $k^2\zeta_k^2$ term is suppressed and the scalar sector becomes strongly coupled unless higher-spatial-derivative operators are included \cite{Arkani-Hamed:2003pdi}. A controlled effective field theory completion of this type is provided by the scordatura mechanism \cite{Motohashi:2019ymr,Gorji:2020bfl,Gorji:2021isn}, i.e.\ a small detuning from exact degeneracy that generates a higher-spatial-derivative operator $\alpha_{\rm sc}(\partial^2\zeta)^2/a^4$ in the quadratic action (equivalently, an $\alpha_{\rm sc}(k^4/a^4)\zeta_k^2$ term in Fourier space). In the standard scordatura setup (with $\beta=0$), this contribution dominates at sufficiently large $k/a$ and yields a ghost-condensate-like UV scaling $\omega \propto \sqrt{\alpha_{\rm sc}c_s^2/\epsilon}(k/a)^2$. By contrast, when $\beta\neq 0$ the additional high-momentum kinetic correction $\propto \beta k^2\dot{\zeta}_k^2/(a^2H^2)$ alters the UV scaling once both effects become relevant, leading instead to a linear dispersion $\omega^2\simeq c_{\rm UV}^2 k^2/a^2$ with $c_{\rm UV}^2\propto (\alpha_{\rm sc}/\beta)$.

For tensor perturbations, up to quadratic order, one has
\begin{align}\label{eq:action-S2T}
    S^{(2)}_{\rm T}  &=  \frac{1}{8} \Mpl^2 \int \D{t}\D^3x {\bar N} a^3 \Bigl[ \dot{\gamma}^{ij} \dot{\gamma}_{ij} -   \frac{1}{a^2} \partial_k\gamma^{ij} \partial^k\gamma_{ij}\Bigr]\;.
\end{align}
We see that there are no modifications in the pure gravity sector compared with general relativity. This is not surprising: indeed, the action~\eqref{eq:action} does not contain any non-minimal coupling between the Ricci scalar and the scalar field. Moreover, by construction, the function $P$ does not include Christoffel symbols. 

The results \eqref{eq:action-S2} and \eqref{eq:action-S2T} show that, on the cosmological background, the theory propagates three healthy modes in the parameter region $\epsilon>0,\; c_s^2 > 0\,$ and $\beta \geq 0$.
Taken together with the Hamiltonian analysis in Sec.~\ref{sec:dof}, which implies that the theory can propagate at most three physical degrees of freedom in unitary gauge, the perturbative result shows that this bound is saturated: the model indeed carries exactly three propagating modes, namely two tensor polarizations and a single scalar.
 
It is worth emphasizing that, although the higher-derivative invariants vanish on a homogeneous FLRW background and therefore affect the dynamics only at the level of perturbations, this is no longer true for less symmetric configurations. In particular, we have explicitly verified that, in a spherically symmetric spacetime, the higher-derivative terms can already contribute at the background level.

\section{Summary and conclusions}
\label{sec:Summary}

In this work we introduced a covariant and gauge-independent construction of scalar-tensor theories endowed with a preferred foliation defined by the timelike gradient of the scalar field, $\nabla_\mu\phi$. Our approach provides a systematic way to build diffeomorphism-invariant operators that are intrinsically spatial on the hypersurfaces $\phi=\mathrm{const}$, organized as a compact basis of independent invariants up to four derivatives of $\phi$. At this derivative order, we also identified the first nontrivial parity-odd pseudoscalar in the scalar-gradient sector and clarified its relation to the parity-even basis through a dimension-dependent identity.

After minimally coupling this sector to gravity, we studied the dynamical content of the resulting class of theories. The Hamiltonian analysis in unitary gauge shows that the theory propagates at most three degrees of freedom. We then explicitly confirmed this expectation by analyzing linear perturbations around a spatially flat FLRW background: all higher-derivative invariants vanish on the homogeneous background, so their first effects arise in perturbations, and the quadratic action exhibits two tensor modes with the standard dispersion relation together with a single scalar mode. In particular, the leading higher-derivative correction at quadratic order is controlled by terms involving second derivatives of the scalar field, while the parity-odd term enters only at much higher order and therefore does not affect linear cosmological perturbations.

Our construction also clarifies the relation to existing degenerate frameworks. DHOST theories are defined as covariant higher-derivative scalar-tensor theories whose Lagrangians satisfy degeneracy conditions ensuring that only one scalar degree of freedom propagates in addition to the two tensor modes. U-DHOST theories enlarge this landscape by requiring degeneracy only in unitary gauge formulation: a covariant rewriting may display higher derivatives and an apparent extra mode, but this mode is non-propagating (``shadowy'') rather than dynamical when $\nabla_\mu\phi$ is timelike. In this perspective, our operator basis goes beyond DHOST already at the level of admissible covariant invariants, and it provides a nonlinear covariant extension of U-DHOST by allowing arbitrary functions of the foliation-spatial building blocks (including nonlinear dependence on $B$, defined in Eq.~\eqref{eq:B-def}, and its higher-derivative descendants), while maintaining a controlled degree-of-freedom count. It would be interesting to study the properties of these theories when $\phi$ is no longer timelike. One possible approach would be to first construct spherically symmetric solutions in which $\phi$ is spacelike, and then to study perturbations around such backgrounds. The dynamics of these perturbations would shed light on the properties of the propagating degrees of freedom.

Recent developments have pushed U-DHOST constructions further, including systematic treatments of healthy scalar-tensor theories with genuine third derivatives of $\phi$ \cite{Michiwaki:2026xru}. Our covariant construction provides a convenient starting point for systematic model building and for future extensions, including the incorporation of non-minimal curvature couplings and a broader investigation of phenomenology and stability in less symmetric backgrounds. It is known that curvature-dependent operators can play an essential role in controlling the dynamics around general relativity solutions, for example, by avoiding strong-coupling issues in certain stealth or timelike-scalar backgrounds \cite{Motohashi:2019ymr,Gorji:2020bfl,Gorji:2021isn,Aoki:2021wew,DeFelice:2022xvq,DeFelice:2022qaz,Aoki:2023bmz,Mukohyama:2025owu}. Extending our covariant basis to include non-minimal couplings to curvature is therefore an important direction for future work.

Finally, it would be interesting to study whether conformal-disformal transformations can be extended, in the spirit of \cite{Babichev:2019twf,Babichev:2021bim,Takahashi:2021ttd,Babichev:2024eoh}, to these new theories. In particular, one may ask whether suitably generalized conformal-disformal transformations can be defined so that  this new class of theories is stable under these transformations, and whether corresponding conformal-disformal equivalence classes can be constructed. This could prove valuable for achieving a deeper understanding of these theories and their coupling to matter.

\subsubsection*{Acknowledgements}
We would like to thank the Institut Pascal, where this work was initiated during a workshop organized in collaboration between IBS and Université Paris-Saclay.
MAG and PP were financed by the Institute for Basic Science under the project code IBS-R018-D3. The work of KN is partially supported by ANR grant StronG (No. ANR-22-CE31-0015-01).

\appendix

\section*{Appendix: Independence of the basis}
\label{app:Independence of the Basis}
\numberwithin{equation}{section}

In $1+3$ spacetime dimensions, any totally antisymmetric tensor with four purely spatial indices vanishes identically. Hence, consider the contraction
\begin{align*}
    \epsilon^{\mu\nu\rho\sigma}\, D_\mu X\, D_\nu B\, D_\rho C^{[1]}\, D_\sigma C^{[2]} \equiv 0 \, .
\end{align*}
Squaring this expression yields the following dimension-dependent identity:
\begin{align*}
   \Big( \epsilon^{\mu\nu\rho\sigma}\, D_\mu X\, D_\nu B\, D_\rho C^{[1]}\, D_\sigma C^{[2]}  \Big)^2
   = -\,\frac{1}{X^{4}}\,
    \begin{vmatrix}
    B & C^{[1]} & D^{[1]} & E^{[1]} \\
    C^{[1]} & C^{[2]} & D^{[2]} & E^{[2]} \\
    D^{[1]} & D^{[2]} & D^{[3]} & E^{[3]} \\
    E^{[1]} & E^{[2]} & E^{[3]} & E^{[4]}
    \end{vmatrix}
    \equiv 0 \, .
\end{align*}
Therefore, the structure associated with $E^{[4]}$ is redundant whenever
\begin{align*}
    \Delta_3 \equiv 
    \begin{vmatrix}
    B & C^{[1]} & D^{[1]} \\
    C^{[1]} & C^{[2]} & D^{[2]} \\
    D^{[1]} & D^{[2]} & D^{[3]}
    \end{vmatrix}
    \neq 0 \, .
\end{align*}
Indeed, on the generic nondegenerate branch where this $3\times 3$ minor is non-vanishing, the vanishing of the $4\times 4$ Gram determinant implies that $E^{[4]}$ can be expressed linearly in terms of the remaining invariants.

We now prove that, on the generic nondegenerate branch, the parity-even basis \eqref{eq:basis} is algebraically independent, whereas \(D^{[\mathrm{odd}]}\) supplies the additional parity-odd information, namely the orientation of the field configuration, which is not determined by the parity-even sector. These results together justify the general form of the Lagrangian written in \eqref{eq:action}.
The proof proceeds by contradiction. Assume that there exists a nontrivial algebraic relation among the basis elements that is identically satisfied for arbitrary field configurations, whether on shell or off shell. Then, in particular, such a relation must also hold when evaluated on any specific family of configurations. To exclude this possibility, we consider the following field configuration:
\begin{align*}
    g_{\mu\nu} &= \mathrm{diag}(-N_0^2,1,1,1) \,, \quad
    \phi = t \,,
\end{align*}
where
\begin{align*}
    N_0^{-2} = a_0 + p\, a_x x + a_{xx} x^2 + \sqrt{a_{xy}}\, x y + a_{yy} y^2 + p\, \sum_{i \leq j \leq k}^{x,y,z} a_{x^i x^j x^k} x^i x^j x^k \,,\quad a_0>0\,,\quad a_{xy}>0\,,
\end{align*}
and, for definiteness, we choose a generic configuration with $a_x>0$ and with the relevant cubic coefficients positive. Here, $p = \pm 1$ encodes the parity of the field configuration. 

For convenience, we have written the term proportional to $x y$ as $\sqrt{a_{xy}}\, x y$. 
This choice is motivated by the fact that $C^{[1]}$ depends only on the square of the coefficient in front of $x y$. 
Hence, this parametrization allows us to avoid unnecessary redundancy.  The explicit expressions for $D^{[3]}$, $E^{[1]}$, $E^{[2]}$, and $E^{[3]}$ are rather lengthy; for this reason, we provide them in a separate Wolfram file, see the ancillary Mathematica file available at~[\href{https://github.com/Furton/Expressions-for-D-and-E}{link}]. This choice of field configuration in the limit $\vec{x} \to 0$ leads to the following system of equations:
\begin{align*}
    X &= - a_{0}{}\,, \quad B = - a_{0}{} a_{x}{}^2\,, \quad C^{[1]} = - a_{0}{} a_{x}{}^4 - 4 a_{0}{}^2 a_{x}{}^2 a_{xx}{}\,,\\
    C^{[2]} &= 
    - a_{0}{} a_{x}{}^6 - 8 a_{0}{}^2 a_{x}{}^4 a_{xx}{} - 16 a_{0}{}^3 a_{x}{}^2 a_{xx}{}^2 - 4 a_{0}{}^3 a_{x}{}^2 a_{xy}{}\,,\\
    D^{[1]} &= 
    - a_{0}{} a_{x}{}^6 - 16 a_{0}{}^2 a_{x}{}^4 a_{xx}{} - 16 a_{0}{}^3 a_{x}{}^2 a_{xx}{}^2 - 12 a_{0}{}^3 a_{x}{}^3 a_{xxx}{} - 4 a_{0}{}^3 a_{x}{}^2 a_{xy}{}\,,\\
    D^{[2]} &= 
    - a_{0}{} a_{x}{}^8 - 20 a_{0}{}^2 a_{x}{}^6 a_{xx}{} - 80 a_{0}{}^3 a_{x}{}^4 a_{xx}{}^2 - 64 a_{0}{}^4 a_{x}{}^2 a_{xx}{}^3 - 12 a_{0}{}^3 a_{x}{}^5 a_{xxx}{} 
    \\
    &- 48 a_{0}{}^4 a_{x}{}^3 a_{xx}{} a_{xxx}{} - 8 a_{0}{}^4 a_{x}{}^3 a_{xxy}{} a_{xy}{}^{1/2} - 12 a_{0}{}^3 a_{x}{}^4 a_{xy}{} - 32 a_{0}{}^4 a_{x}{}^2 a_{xx}{} a_{xy}{} 
    \\
    &- 16 a_{0}{}^4 a_{x}{}^2 a_{xy}{} a_{yy}{}\,,\\    D^{[\mathrm{odd}]} &= 8  a_{0}{}^{3} a_{x}{}^4 a_{xxz}{} a_{xy}{}^{1/2} p\,.
       \end{align*}
It is instructive to rewrite the system in a form in which each basis variable 
$X$, $B$, $C^{[1,2]}$, $D^{[1,2,3]}$, and $E^{[1,2,3]}$ depends on a distinct and 
independent parameter of the field configuration, namely 
$a_{x^i}$, $a_{x^i x^j}$, and $a_{x^i x^j x^k}$. In particular, we write
\begin{align*}
  X &= -a_{0}\,, 
  \qquad 
  B = a_{x}^{2}\, X\,, \\
  C^{[1]}{} &= \frac{B^2}{X} - 4 a_{xx}{} B X \,,\quad  C^{[2]}{} = \frac{C^{[1]}{}^2 X^2 + 4 a_{xy}{} B^2 X^4}{B X^2}\,,\\
 D^{[1]}{}
 &= C^{[2]}{} + \frac{2 (- B^3 + B C^{[1]}{} X + 6 a_{xxx}{}(B^3 X^7)^{1/2})}{X^2}  \,,\\
 D^{[2]}{} -  E^{[1]}{}
 &= \frac{C^{[1]}{} (C^{[2]}{} -  D^{[1]}{})}{B} + \frac{2 C^{[1]}{}^2}{X} -  \frac{2 B C^{[2]}{}}{X} + 4 a_{xxy}{} \Big(  B X^{3}[  B C^{[2]}{} - C^{[1]}{}^2]\Big)^{1/2} \,,\\
D^{[2]}{} -  \tfrac{1}{2} E^{[1]}{} 
&= - \frac{B^3 C^{[2]}{} + C^{[1]}{}^3 X - 2 B C^{[1]}{} X (C^{[2]}{} + 2 a_{yy}{} C^{[1]}{} X) -  B^2 (C^{[1]}{}^2 - 4 a_{yy}{} C^{[2]}{} X^2)}{2 B^2 X}\,.\\
\end{align*}
Thus, within the chosen field configuration, each parity-even basis variable is controlled by its own independent parameter. By contrast, the parity-odd invariant \(D^{[\mathrm{odd}]}\) depends on the additional discrete parameter \(p\), which does not enter any parity-even basis element. Hence, the parity-even sector is algebraically independent, while \(D^{[\mathrm{odd}]}\) carries the extra information associated with the orientation of the configuration. Since all other basis variables are insensitive to \(p\), \(D^{[\mathrm{odd}]}\) is not redundant; rather, it uniquely encodes the parity of the configuration, namely the sign of \(p\).

Finally, let us show that all remaining pseudoscalars are redundant.
Indeed, the square of any parity-odd pseudoscalar can be written in terms of Gram determinants built solely from the parity-even basis variables. As an illustrative example, consider
\[
n_\mu \epsilon^{\mu\nu\rho\sigma}\,D_\nu X\,D_\rho B\,D_\sigma C^{[2]} \, .
\]
Its square is given by
\begin{align*}
\Big(n_\mu \epsilon^{\mu\nu\rho\sigma}\,D_\nu X\,D_\rho B\,D_\sigma C^{[2]}\Big)^2
= \frac{\Delta_E^{\,2}}{X^3\,\Delta_3}\,,
\end{align*}
where
\begin{align*}
\Delta_E \equiv
\begin{vmatrix}
B & C^{[1]} & D^{[1]}\\
C^{[1]} & C^{[2]} & D^{[2]}\\
E^{[1]} & E^{[2]} & E^{[3]}
\end{vmatrix},
\end{align*}
provided that $\Delta_3\neq 0$. The same conclusion holds, on the generic nondegenerate branch, for any other parity-odd pseudoscalar. Therefore, every such pseudoscalar is determined, up to an overall sign, by the parity-even basis variables. Since this sign is already encoded in $D^{[\mathrm{odd}]}$, no additional independent parity-odd pseudoscalars beyond \eqref{def:Dodd} arise on this branch. This completes the proof.

\bibliographystyle{JHEPmod}
\bibliography{refs}
	
\end{document}